\documentclass[12pt]{article}
\usepackage{graphicx}
\usepackage[superscript]{cite}

\newcommand{\dir}{Figs}

\newcommand{\ud}{\mbox{d}}

\newcommand{\vq}{\mbox{$\mathbf{q}$}}

\newcommand{\kB}{\mbox{${k_{_B}}$}}
\newcommand{\ie}{\mbox{\em i.\ e., }}
\newcommand{\eg}{\mbox{\em e.\ g., }}
\newcommand{\etal}{\mbox{\em et al. }}

\begin{document}

\hspace*{1.5cm}
\begin{minipage}{13cm}
{\Large \bf Toy amphiphiles on the computer: What can we 
learn from generic models?}

\vspace{2\baselineskip}

{\em Friederike Schmid,${}^{*1}$}

\vspace{\baselineskip} 

${}^1$Physics Department, University of Bielefeld, Universit\"atsstrasse 25, 
D-33615 Bielefeld, Germany 

\vspace{\baselineskip}

{\bf Summary:}

Generic coarse-grained models are designed such that they are 
(i) simple and (ii) computationally efficient. They do not aim at
representing particular materials, but classes of materials, hence 
they can offer insight into universal properties.
Here we review generic models for amphiphilic 
molecules and discuss applications in studies of self-assembling 
nanostructures and the local structure of bilayer membranes, \ie their 
phases and their interactions with nanosized inclusions. Special
attention is given to the comparison of simulations with elastic continuum 
models, which are, in some sense, generic models on a 
higher coarse-graining level. In many cases, it is possible to
bridge quantitatively between generic particle models and 
continuum models, hence multiscale modeling works on principle. 
On the other side, generic simulations can help to interpret
experiments by providing information that is not accessible otherwise.
\vspace{\baselineskip}


\end{minipage}

\vspace{1.5\baselineskip}


{\large \bf 1. Introduction}

\begin{figure}[t]
\centerline{
\includegraphics[width=0.48\textwidth]{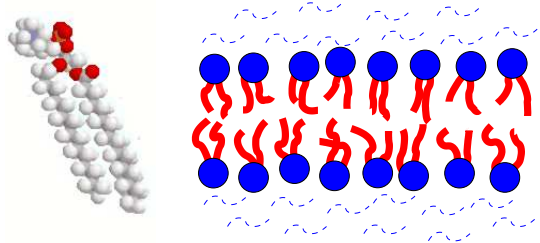}  
}
\caption{\label{fig:lipids} 
Left: Example of a lipid molecule (DPPC).
Right: Schematic sketch of a lipid bilayer.
}
\end{figure}

Amphiphiles are key constituents of living matter and of many technologically
important substances and materials\cite{book_GS,book_I,book_GBR}. 
In the literal sense, the term "amphiphile" describes a chemical compound 
containing hydrophilic as well as hydrophobic parts. More generally,
it is used for compounds where chemically incompatible units are permanently 
linked together, such as block copolymers. Amphiphiles are highly surface 
active, {\em i.e.}, they segregate to interfaces and surfaces and alter
their interfacial properties. At high amphiphile concentrations, amphiphilic
substances have a propensity to ``microphase separate'', \ie to develop complex
structures on the mesoscale that contain many internal interfaces. As an example,
consider lipids, a particularly prominent class of amphiphilic molecules. 
They are made of one polar (hydrophilic) head group connected to two or more 
nonpolar (hydrophobic) tails (see Fig.\ \ref{fig:lipids} a). 
In water solution, they self-assemble into a variety of nanostructures, 
\eg micelles, vesicles, sponges, or lamellae.  
Many of these structures share as common element the lipid bilayer, a stack
of two opposing lipid monolayer sheets, where the lipids are arranged such 
that the hydrophobic tails are shielded from the water by the hydrophobic
heads (Fig.\ \ref{fig:lipids} b)). Lipid bilayers play a central role in
biophysics, since they provide the basic frame for biomembranes\cite{book_G,book_LS}.

Amphiphilic systems exhibit two particularly striking features,
which have attracted theoretical interest for many decades: 
The {\em self-assembly} of molecules into complex structures, often associated
with mesophase formation, and very peculiar {\em interfacial properties}, both
for interfaces that separate macrophases and for interfaces that are part of 
a mesophase (as in the case of membranes). Traditionally, theorists have
taken different routes to describe these different aspects of amphiphilic 
systems. While the self-assembly and mesophase formation is usually treated 
within approaches based on bulk thermodynamics, \eg packing 
arguments\cite{book_I}, or (bulk) field theories of varying 
complexity\cite{book_GS,review_M}, the interfacial aspects have often been 
discussed in terms of ``effective interface'' theories, where the essential 
degrees of freedom are assumed to be localized on effectively two-dimensional 
manifolds, the interfaces\cite{book_S}. The most famous example of an 
effective interface Hamiltonian is the ``Helfrich Hamiltonian'', introduced by 
Helfrich\cite{H73} in 1973, which relates the free energy of bilayer 
membranes to the invariants of their local curvature tensor, 
\begin{equation}
\label{eq:helfrich}
{\cal H} = \int \ud A \big\{ \sigma + 2 \kappa (H - c_0)^2
+ \frac{1}{2} \bar{\kappa} K \big\}.
\end{equation}
Here $H = (c_1+c_2)/2$ (mean curvature) and $K=c_1 c_2$ (Gaussian curvature),
with the principal curvatures $c_1$ and $c_2$ (\ie the Eigenvalues of the 
curvature tensor). The Helfrich Hamiltonian involves four phenomenological
parameters, the surface tension $\sigma$, the bending modulus $\kappa$, 
the saddle-splay modulus $\bar{\kappa}$, and the spontaneous curvature $c_0$,
which depend on the molecular structure of the membrane. Bulk membranes
at equilibrium should be tensionless ($\sigma = 0$), and symmetric bilayers
should not have a spontaneous curvature ($c_0=0$). Moreover, the contribution
of the Gaussian curvature is a constant for closed surfaces with fixed
topology. This leaves one with only one parameter $\kappa$ in the simplest case. 
The Helfrich Hamiltonian is often used as starting point for more complex elastic
theories of monolayers and bilayers, which incorporate additional factors 
such as the membrane thickness, internal degrees of freedom, interlayer
coupling etc.

The third route to studying amphiphilic systems is of course the use of computer 
simulations. Since simulations in full atomic detail are very expensive 
and the system sizes accessible for such simulations are still limited, 
coarse-grained models are widely applied to investigate various aspects of 
amphiphilic systems. Two different coarse-graining philosophies
have been pursued, {\em systematic} coarse-graining and {\em generic} 
coarse-graining. In the first approach, coarse-grained models are derived
more or less systematically from atomistic models. The vision is to develop 
strategies for constructing whole hierarchies of coarse-grained models
for specific materials, which can then be used to make quantitative 
predictions of material properties\cite{V88,SP98,F02,IV05}. In the second 
coarse-graining line, idealized models are developed which incorporate 
only properties of amphiphiles that are deemed essential for their particular 
behavior.  Such ``generic'' coarse-grained models are less quantitative 
than systematic coarse-grained models, but they give insight into 
basic physical mechanisms which are responsible for the peculiar 
properties of amphiphiles, and they can make predictions for whole
classes of materials. Generic coarse-graining has a long-standing 
tradition in the theory of amphiphilic systems; the earliest model,
the Wheeler-Widom model (a simple Ising-type lattice model) dates
back to 1968\cite{WW68}. Nowadays, a whole zoo of lattice
and off-lattice models has been proposed and used to study various 
aspects of amphiphilic systems. Discussing them all is far beyond the 
scope of this article. A number of review articles have appeared
that provide an overview over coarse-grained models for self-assembling 
amphiphilic systems in general\cite{book_GS,S00,S06,K07} or for membranes in 
particular\cite{MKS06,VSS06,BLB06,SL06,S08}. Here, we shall solely give examples, 
mainly from our own work.

In the next section, we discuss equilibrium and dynamical aspects of
amphiphile self-assembly on mesoscopic scales -- the traditional realm
of generic amphiphile models. In section three, we examine
bilayer membranes on a more local scale, most notably, internal membrane 
phase transitions. Even on this scale, generic models can provide 
valuable insights into basic processes that govern membrane (bio)physics.
We conclude with a brief outlook in section four.

\vspace*{\baselineskip}

{\large \bf 2. Self-assembly and mesoscale structure}

\vspace*{0.5\baselineskip}

{\bf Modeling}

Most generic models that have been used to simulate amphiphile self-assembly
and mesostructure formation in amphiphilic systems belong to one of
three categories: Lattice spin models\cite{HK88,GD91,FS94,MS94,LS98}
particle models\cite{LSD85,L96,LBS96,SSR90,SSR91,KEH94,GL98}, 
and field-based models -- the latter may be purely phenomenological\cite{book_GS}
or derived from (coarse-grained) molecular models\cite{MKS06}. 
Here we will discuss particle- or field-based models with an underlying 
particle picture. These models have in common that they identify as 
crucial factor the amphiphilic character of the molecules,
\ie the fact that they are made of two chemically incompatible blocks.

The central idea of generic coarse-graining is to restrict the description
of a system to bare essentials, \ie to simplify as much as possible. 
One particularly simple computer model for self-assembling amphiphiles 
has been proposed in 1990 by Smit \etal \cite{SSR90,SSR91} and has since 
become an archetype off-lattice model for amphiphilic systems. 
(An analogous, similarly successful chain model on a lattice had 
been introduced five years earlier by Larson\cite{LSD85}.)
The basic elements are beads, which may have one of two types, 
{\em w} (water-like) or {\em o} (oil-like). The interactions between 
beads are chosen such that {\em w} beads and {\em o} tend to segregate. 
Amphiphilic molecules are represented by chains of {\em w} and {\em o} beads. 
This model produces micelles and bilayers -- indicating that the
amphiphilic character of molecules is indeed sufficient to drive
self-assembly.

A second more practical requirement for generic computer models 
is computational efficiency. Since the details of the potentials 
are usually not in the focus of interest, they may as well be 
chosen such that computer simulations are cheap. In recent years, a
number of models that are very interesting from this point of view have 
been developed at the Max-Planck Institute for polymer science in Mainz.
Soddemann \etal \cite{SBK01} have designed a Smit-type model which
is optimized such that the basic bead structure is largely 
that of a simple hard sphere liquid, independent of the overlying
oil/water/amphiphile ''colouring''.  To this end, the range of 
attractive interactions and the molecular bond length are chosen 
such that they match the interparticle distance in the liquid. 
Length scale frustrations that slow down the simulations are 
thus avoided; moreover, semi-grandcanonical Monte Carlo identity 
switches are facilitated considerably. 
The Soddemann model and variants thereof have been used to study 
structure formation in amphiphilic systems under equilibrium and 
nonequilibrium conditions\cite{GKS02a,GKS02b,SAG04,LMS03,LMS04,LMS05,
GC05,G06,FDM06}.
Another, even cheaper model that is specifically designed to study
large scale properties of bilayers has recently been proposed by Cooke 
\etal\cite{CKD05}. Amphiphiles are represented by chains of three
beads (one {\em w}, two {\em o}), and because of a smart choice of
potentials, the solvent (water) can be omitted altogether.
From the point of view of computational efficiency, implicit solvent 
models are of course particularly appealing, and numerous studies had
shown that they are suitable to study self-assembled structures in binary 
amphiphile/water systems\cite{GSH97,GSH98,BMC98,NT01,F03,BPB05,WF05}.
As an alternative to implicit solvent models, we have proposed a 
``phantom'' solvent model, where the solvent only interacts with
lipids\cite{LS05,SDL07}. In Monte Carlo simulations, which is only slightly 
more expensive.

The most optimized particle-based models for amphiphilic systems, 
represent amphiphilic molecules by just a few elementary units 
($\sim$ 2-4 beads). Such descriptions are highly successful for short-chain 
amphiphiles. In the case of (co)polymeric amphiphiles, however, the 
chain character of the molecules may become important, and 
coarse-grained models should incorporate this aspect. Therefore,
particle-based simulations of polymeric amphiphiles have
often resorted to using established coarse-grained polymer models, 
such as the bond-fluctuation model\cite{CK88},
or similarly popular off-lattice models\cite{GLK96}.
Unfortunately, simulations of long-chain polymeric systems at high
densities are expensive even with the most optimized particle-based 
polymer models. This motivates the use of field-based models that 
propagate density fields (or related fields) instead of particles. 
Such models can be derived more or less systematically 
from a chain molecule picture {\em via} a route borrowed from
one of the most successful polymer mean-field theories,
the self-consistent field (SCF) theory\cite{HT71a,HT71b,S98}.
In the SCF approximation, systems of interacting chains
are replaced by an ensemble of independent chains in an 
inhomogeneous field, which is determined self-consistently.
Already early applications of this approach have dealt with
amphiphilic systems, \eg it has been used to study micelle 
structures\cite{S93} or microphase separation in block copolymer 
melts \cite{WN85,MS94b}. Compared to particle-based simulations
of block copolymer/homopolymer mixtures, SCF theories can
reproduce local structures at a {\em quantitative} 
level\cite{WSB96,WSMB99b}, provided that one identifies
the correct ``intrinsic coarse-graining length''\cite{WSMB99a}.
More recently, methods to include dynamics\cite{F93,HF97,MF97,MZF98} 
and fluctuations beyond mean-field\cite{GF01,FGD02,DGF03}
have been developed. Together, these methods constitute 
the new class of ''molecular field-based'' (or ''field-theoretical'') 
simulation methods for polymeric systems. The derivation of 
field-based simulation models is complicated and technical and 
shall not be presented here. The interested reader is referred 
to recent reviews\cite{MS05,book_F}.

After this brief and highly incomplete review of approaches
to modeling amphiphile self-assembly on ``large'' scales
(\ie nm to $\mu$m range) at a generic level, we proceed to 
presenting actual simulation studies from our group. We
shall ask two questions: (i) To which extent can elastic
theories such as the Helfrich theory (\ref{eq:helfrich}) 
describe the properties of self-assembled amphiphilic systems,
and (ii) how do amphiphiles self-assemble in practice, \ie, 
which are the kinetic pathways of self-assembly?

\vspace*{\baselineskip}

{\bf Fluctuating mesoscale structures versus elastic theory}

\begin{figure}[t]
\centerline{
\includegraphics[width=0.5\textwidth]{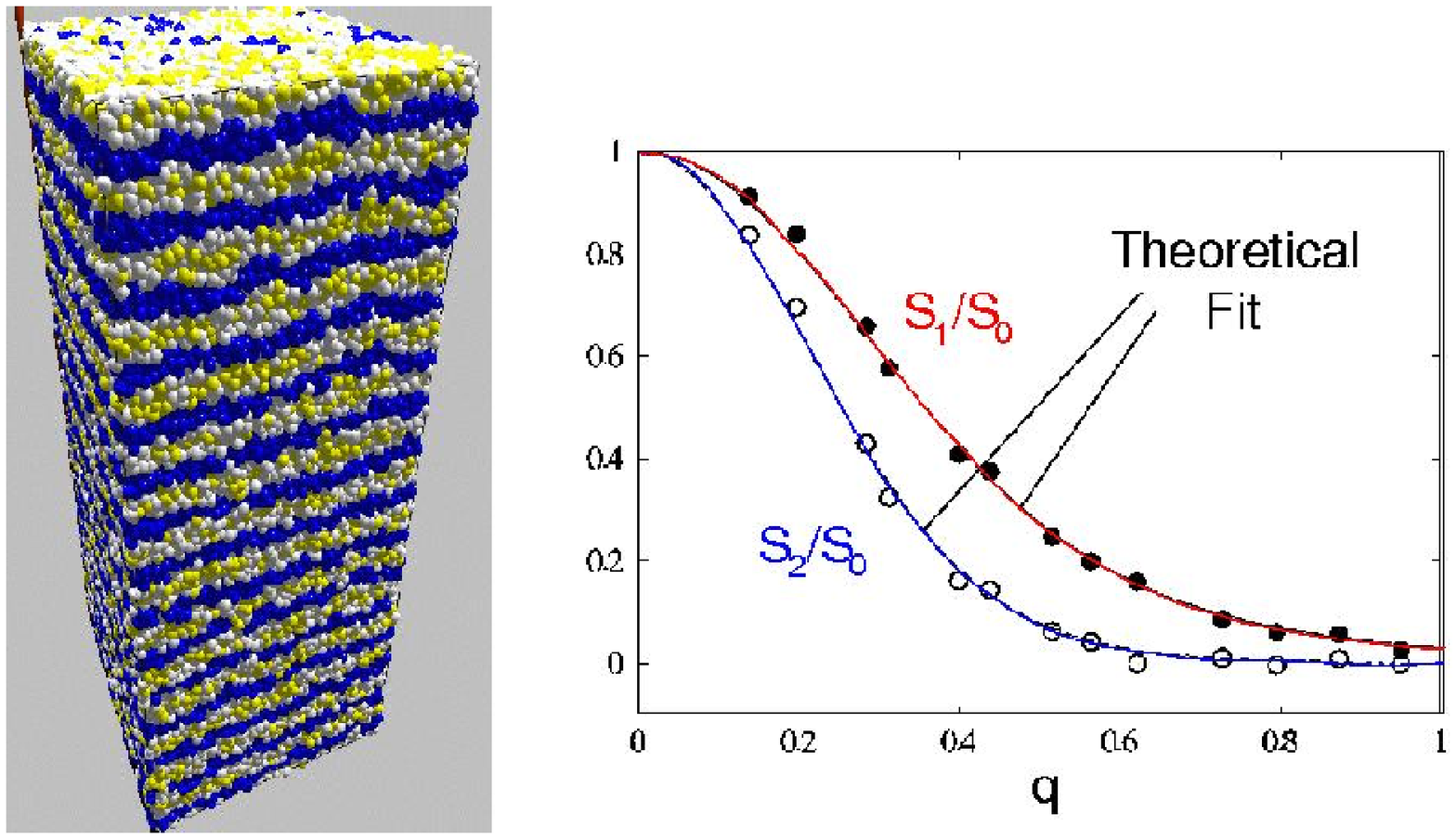}  
}
\caption{
\label{fig:loison} 
Left: Snapshot of a lamellar stack from a coarse-grained
simulation of $w_20_2$ amphiphiles with 20 \% $o$-solvent.
Right: Transmembrane structure factor for this system vs.\
in-plane wavevector $q$ (arbitrary units), compared with
elastic theory. See text for explanation.
(from Ref.\ \citeonline{LMS03}).
}
\end{figure}

In order to address the first question mentioned above, 
we have used a generic particle-based model which exhibits
swollen lamellar phases, and compared the properties
of the structures observed in the simulations with
the predictions of appropriate continuum theories. Specifically,
we have used a variant of the Soddemann model\cite{SBK01},
where amphiphilic molecules are represented by
$w_2o_2$ tetramers and dissolved in $w$ solvent.
At sufficiently low temperatures, the systems develops 
a lamellar phase. We found that the lamellae can incorporate 
roughly 40 volume \% solvent without being destroyed.
The simulations were carried out at 20 \% solvent. A snapshot
is shown in Fig.\ \ref{fig:loison} (left). The system has
a relatively high degree of order, but thermal fluctuations 
and membrane defects are still prominent.

According to the simplest elastic model\cite{LSB95}, 
the free energy of a lamellar stack is given by
\begin{equation}
\label{eq:lei}
{\cal H} = \sum_n \int \ud A
\big\{
\frac{\kappa}{2} (\Delta u_n)^2
+ \frac{B}{2} (u_{n+1}-u_n)^2
\big\},
\end{equation}
where $u_n(x,y)$ denotes the local deviation of the height of
the $n$th lamella from its average value, $\kappa$ is the
bilayer bending modulus introduced earlier (\ref{eq:helfrich}),
and $B$ the compressibility of the stack. This free energy
determines the amplitude of thermal fluctuations of the
lamellar position $u_n$. As an example, we consider the 
``transmembrane structure factor'' 
$S_n(q_x,q_y) = \langle u_n(\vq) u_0(\vq)^* \rangle$, which
can be calculated analytically\cite{LMS03} ($u_n(\vq)$ is 
the two-dimensional Fourier transform of $u_n(x,y)$). 
Fig.\ \ref{fig:loison} (middle) shows a comparison 
between theory and simulation, with only one fit parameter
$\xi = (\kappa/B)^{1/4}$. The agreement is excellent\cite{LMS03}.

A similarly simple free energy model for pore defects in membranes, 
where the statistics of pore shapes is taken to depend only on the
line tension, was found to perform equally well in comparison
with the simulations\cite{LMS04}. Even the behavior of long
polymers inserted in the membrane can be understood by scaling 
arguments that are based on the elastic theory of membrane stacks 
(hydrophilic polymer collapse and create exactly one pore)
\cite{LMS05}.

We conclude that our generic simulations confirm the 
validity of elastic models for the description of fluctuating
amphiphilic bilayer systems on the mesoscale. We shall see
later that they even perform surprisingly well on the scale 
of the membrane thickness. 

\vspace*{\baselineskip}

{\bf Kinetics of self-assembly}

Our second question relates to dynamical aspects of self-assembly.
Here, we were particularly interested in the pathways leading
to the formation of amphiphilic copolymeric vesicles or other 
self-assembled copolymeric nanostructures. The work was motivated
by increasing recent experimental interest in these 
structures\cite{ZE96,ZYE96,SE00,HLS06,JB03,PCC04}. 
Artificial vesicles are expected to have a high potental in nanotechnology 
as microreactors or microcontainers. Since it may take weeks or months 
before vesicle solutions are truly equilibrated, the vesicles observed
in experiments are often nonequilibrium structures which depend 
on the history of the system, \ie on the kinetic pathways of
self-assembly. These pathways are hard to unravel experimentally,
hence computer simulations can provide useful insight.

Previous simulation studies of vesicle self-assembly have revealed
one possible pathway to vesicle formation (hereafter referred to
as pathway I): In a first step, small micelles form; then, the 
micelles coalesce to disks, \ie small bilayer fragments; finally, 
the disks curve around and close up to form vesicles. These simulations 
focussed on short-chain amphiphiles\cite{LBS96,NT01b,YMH02,VMM04,SZ05,SZ07}. 
In copolymeric amphiphiles, the local driving forces for segregation are 
weaker and the diffusion is slower. The question was whether this has 
an effect on the pathway and the final self-assembled nanostructures.

To study this problem, we used a coarse-grained field-based model,
the ''external potential dynamics'' (EPD) method developed by 
Maurits and Fraaije in 1997 \cite{MF97}. We study an underlying 
particle model where amphiphiles are represented by linear strings 
(``Gaussian chains'') with a short hydrophilic block $A$
attached to a longer hydrophobic block $B$ (length ratio 2:15),
immersed in a solvent $S$. In EPD, the chains are taken to propagate 
in their surrounding self-consistent field according to 
``Rouse dynamics''\cite{book_DE}, \ie the effect of chain connectivity 
is accounted for, but entanglements and hydrodynamic effects are neglected. 
The system is characterized by three interaction parameters 
$\chi_{AB}$, $\chi_{AS}$, and $\chi_{BS}$, which describe the 
mutual incompatibility of $A$ chain segments, $B$ chain segments, 
and the solvent $S$ (the larger $\chi_{ij}$, the more incompatible $i$ 
and $j$). Of these, the parameter $\chi_{BS}$ turned out to be the most 
influential, hence the other two were kept fixed. Another important 
quantity is the volume fraction $\Phi_p$ of copolymers.

\begin{figure}[t]
\centerline{
\includegraphics[width=0.49\textwidth]{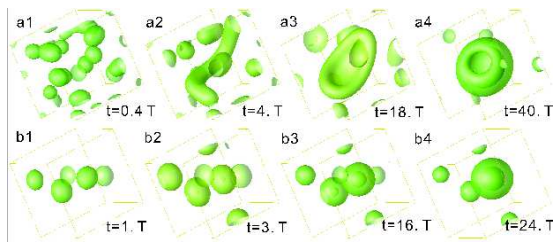} 
}
\caption{\label{fig:he1} 
Pathways of spontaneous vesicle formation in copolymer solutions.
Top: Pathway I observed at a copolymer volume fraction of 20 \%
with (a1) micelle formation, (a2) micelle coalescence, (a3) bilayer
formation, and (a4) bending into vesicle.
Bottom: Pathway II observed at a copolymer volume fraction of 15 \%
with (b1) micelle formation, (b2) micelle growth, (b3) internal
reorganization into semivesicle, and (b4) swelling into vesicle.
From Ref.\ \citeonline{HS08}.
}
\end{figure}

The simulations revealed that spontaneous vesicle formation from
homogeneous solution may in fact proceed {\em via} two distinct pathways
(see Fig.\ \ref{fig:he1}).  At high copolymer concentrations $\Phi_p$, 
we recover the ``traditional'' pathway I described earlier,
including micelle coalescence, sheet formation, and sheet
bending (Fig.\ \ref{fig:he1}, top). At low copolymer concentrations,
a new pathway is observed, which is characterized by growth
processes rather than aggregation (pathway II). Here the first step
is micelles formation as in pathway I, but instead of coalescing,
the micelles then keep growing by incorporating more and more 
copolymers from the solution until the radius of the hydrophobic 
core exceeds the radius of gyration of the $B$ block. Then, 
copolymers start flipping such that the micelle core becomes 
hydrophilic (``semivesicle state''). Finally, solvent diffuses 
inside the core, and the semivesicle swells to form a 
vesicle\cite{HS06}. 

\begin{figure}[t]
\centerline{
\includegraphics[width=0.49\textwidth]{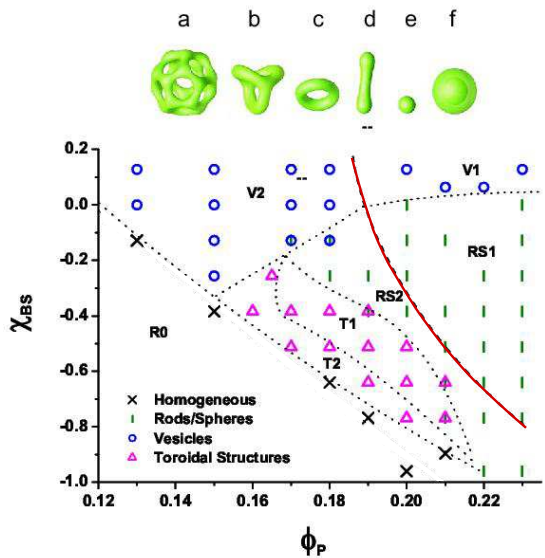}  
}
\caption{\label{fig:he2} 
``Phase diagram'' of final structures after a quench from an
initially homogeneous $A:B$-copolymer solution for a range of
$B$-Solvent interactions $\chi_{BS}$ and copolymer volume fractions
$\Phi_p$. The final structures in the regions V1/V2 correspond to
vesicles (f), RS1/RS2 to rod-sphere mixtures (d,e), T1 to ring 
micelles (c), and T2 to toroidal micelles (b,a). In the region
R0, the solution stays homogeneous. The solid line separates two 
dynamical regions where the structure formation proceeds along
pathway I (regions RS1 and V1) and pathway II (regions RS2,T1,T2,V2). 
The dotted lines are guides for the eye. After Ref.\ \citeonline{HS08}.
}
\end{figure}

This second pathway turns out to be auspicious, as it can 
be exploited to manipulate the sizes and shapes of self-assembled 
nanostructures. For example, the final size distribution
of vesicles can be influenced by mixing seeds into the initial
homogeneous solution\cite{HS06b}. At low copolymer concentration,
the vesicles developing from such seeds may develop hierarchical
multicompartment structures\cite{HS06b}. Even in the absence of 
seeds, micelles with complex toroidal or net-cage structures may
form at low copolymer concentrations close to the CMC (critical 
micelle concentration)\cite{HS08}. A dynamical ``phase diagram''
of final structures after a quench from homogeneous solution
(no seeds) is shown in Fig.\ \ref{fig:he2}.

The simulations covered length scales in the $\mu$m range and
time scales up to almost one second. Such length and time scales
are difficult to access with particle-based models, let alone
atomistic models. They demonstrate the power of field-based
simulation methods to unravel basic mechanisms of self-assembly.
The simulations can not only help to understand experiments,
they may also be used to guide them, \eg in order to produce
certain types of nanostructures in a controlled way.

\vspace*{\baselineskip}

{\large \bf 3. Membrane structure}

\vspace*{0.5\baselineskip}

{\bf Lipid membranes}

In the second part of this paper, we discuss the value of generic
models for studying internal properties of lipid membranes, \ie 
structural properties on length scales of the membrane thickness $d$
or just a few $d$ ($\sim 100 d$). Unlike for mesoscale structures, 
it is far from obvious that generic models are of any use here. 
One might rather suspect that the membrane properties on such 
small scales are dominated by specifics of the molecular structure 
of the constituting lipids. On the other hand, experimental 
bilayer studies for a wide class of lipids indicate that the 
behavior of membranes is to a large extent dictated by universal, 
nonspecific factors\cite{KC94,KC98}. 

At high temperatures, lipid bilayers usually assume a ``fluid''
state which is characterized by a relatively high lipid mobility,
low shear viscosity, and a high degree of disorder in the lipid
tails, the $L_\alpha$ phase. 
Upon decreasing the temperature, one encounters a so-called ``main'' 
transition to a more ordered ``gel'' state with lower mobility. 
The characteristics of the low-temperature phase are mostly 
determined by the geometry of the lipids. For lipids with small head 
group volumes such as, \eg phosphatidylethanolamines, 
chains in the gel state are on average untilted with respect 
to the bilayer normal (the $L_\beta$ phase).
Lipids with larger head groups such as, \eg phosphatidylcholines,
usually assume a gel state where the lipids exhibit collective tilt
(the $L_{\beta'}$ state).  In some cases, where the head-head
attractions are weak, they may also form an untilted $L_{\beta}^{int}$
state where opposing lipid layers are fully interdigitated.
These observations seem to be quite universal, which raises
the hope that the basic characteristics of the main transition
can be reproduced by suitable generic models\cite{KS03,KS05}.

The main transition from the fluid state $L_\alpha$ to the tilted
gel state $L_{\beta'}$ is particularly intriguing, because it actually
proceeds in two steps. There exists an intermediate phase $P_{\beta'}$,
first discovered by Tardieu in 1973\cite{TLR73}, which is characterized
by periodic stripe modulations. This so-called ``ripple'' phase has been 
studied intensely with various methods, \eg calorimetry, atomic force 
microscopy, NMR, and extensive $X$-ray measurements, but nevertheless, it 
has not yet been possible to determine the exact microscopic structure
by experiments. We shall see that generic computer simulations can 
be of use here.

It is clear that the optimized amphiphile models discussed in the previous 
section are too simple to be a good starting point for studies of internal 
membrane transitions. Theoretical mean-field calculations\cite{FS95} indicate 
that the main transition is driven by an interplay between the conformational
entropy of the tails and their tendency to develop nematic order, 
hence a good model should take into account the chain character
of lipids. Our membrane model\cite{SDL07} is based 
on an amphiphile model that has already been used successfully to 
study phase transitions in Langmuir monolayers
\cite{HHB95,HHB96,HH96,SL97,CLS99,CS99,DS01}
The lipids are modeled by semiflexible chains of 6 ``tail'' beads attached 
to one slightly larger ``head'' bead. Tail beads attract each other, 
whereas head beads are purely repulsive. This drives a local segregation 
of heads and tails. In addition, a fluid of ``phantom solvent'' beads
drives the self-assembly of lipids into membranes (see above\cite{LS05}).
  
\begin{figure}[t]
\centerline{
\includegraphics[width=0.45\textwidth]{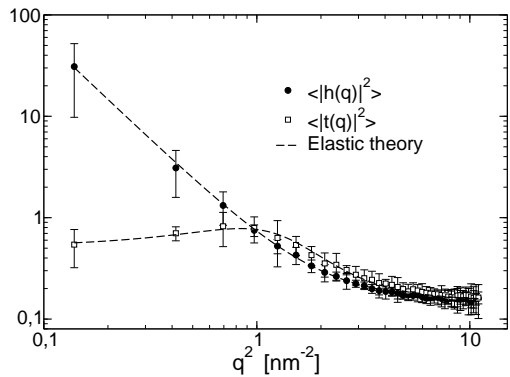}  
}
\caption{\label{fig:fluctuations} Fluctuation spectrum of height and 
thickness fluctuations of a planar membrane, simulated with our
generic membrane model\cite{SDL07}, compared with the elastic
theory of Brannigan and Brown\cite{BB06}.
Here $h(\vq)$ and $t(\vq)$ are the Fourier transforms of the local
mean position and monolayer thickness of the membrane,
$h(x,y) = (z^+(x,y)+z^-(x,y))/2$ and $t(x,y) = (z^+(x,y)-z^-(x,y))/2$, 
where $z^+(x,y)$ and $z^-(x,y)$ are the local positions of 
the opposing head group layers. After Ref.\ \citeonline{WBS09}.
}
\end{figure}

Like real phosphotidylcholine bilayers, the model exhibits a 
fluid $L_\alpha$ phase at high temperatures and a tilted gel $L_{\beta'}$ 
phase at lower temperatures\cite{LS07}. We have examined
in detail the properties of fluid membranes at a temperature slightly
above the main transition. They turn out to be surprisingly `realistic'.
The ratio of area per lipid and squared monolayer thickness roughly 
corresponds to that of DPPC bilayers ($\sim 0.16$). We thus use the properties 
of DPPC bilayers to map our intrinsic model units to SI units
-- more specifically, we match the bilayer thickness and the temperature of 
the main transition to identify the length and energy scale in our system.
The elastic parameters of the membrane can be determined from the stress 
tensor profile across the membrane\cite{book_S} 
and from an analysis of membrane height and thickness fluctuations. 
We use as reference an elastic theory due to 
Brannigan and Brown\cite{BB06}, which treats membranes as a system of two 
coupled elastic monolayer sheets and also accounts for protrusions.
Fig.\ \ref{fig:fluctuations} shows that this theory fits the simulation 
data excellently\cite{WBS09}. The resulting elastic parameters
$\kappa$ (bilayer bending modulus), $k_A$ (area compressibility),
and $c_0$ (spontaneous curvature of the monolayer) are given by
$\kappa \approx 2.2 \cdot 10^{-20}$J, $k_A \approx 130$mN/m, 
and $c_0 \approx -0.08$/nm.
In comparison, experimental values for DPPC are\cite{M06} 
$\kappa \sim 5-20 \cdot 10^{-20}$J and $k_A \sim 230$mN/m, 
and values from fully atomistic simulations are 
$\kappa \sim 4\cdot10^{-20}$J \cite{LE00},
$k_A \sim 300$ mN/m \cite{LE00}, and 
$c_0 \sim (-\!0.02) - (-\!0.05)$/nm \cite{MRY07}. 
Our simple generic membrane model thus not only recovers general 
the elastic behavior of fluid membranes, but the material constants
even have the correct order of magnitude. This suggests that 
the range of elastic constants is to a large extent determined by generic factors.

\begin{figure}[t]
\centerline{
\includegraphics[width=0.3\textwidth]{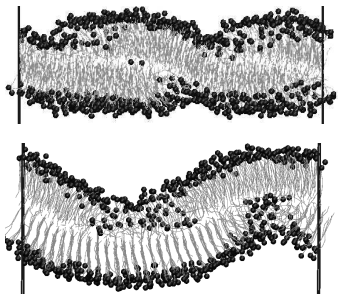}  
}
\caption{\label{fig:ripple} 
Two snapshots of ripple states in our generic membrane
model. Only heads (reduced size) and tail bonds are shown.
Top: Asymmetric ripple state, obtained after cooling rapidly from
the fluid $L_\alpha$ state.
Bottom: Symmetric ripple state, obtained after cooling slowly
from the fluid $L_\alpha$ state.
After Ref.\ \citeonline{LS07}.
}
\end{figure}

These results are already rewarding. The biggest success of our generic model, 
however, is that it recovers the modulated intermediate phase between the 
fluid $L_{\alpha}$ phase and the lower temperature $L_{\beta'}$ phase,
which is observed in real lipid membranes. The intermediate state reproduces 
many features of the experimental $P_{\beta'}$ (ripple) state: There exist 
two modifications, one which is asymmetric and one which is symmetric
with twice the period of the asymmetric structure\cite{TYH98,KTL00} -- both 
structures are observed in simulations (see snapshots in Fig.\ \ref{fig:ripple}).
As in experiments, the asymmetric state is observed after cooling
rapidly from the fluid $L_\alpha$ state or heating from the $L_{\beta'}$ 
state, and the symmetric state is observed after cooling slowly from the 
fluid state. In both ripple states, most of the chains are highly ordered, 
much like in the gel state. The self-diffusion of lipids, on the other hand, 
is significantly higher than in the gel state, and highly anisotropic, 
suggesting that ripple states contain anisotropic ``coexisting'' gel-state 
and and fluid-state domains. Indeed, such ordered and disordered stripes
can clearly be identified in the simulations.

\begin{figure}[t]
\centerline{
\includegraphics[width=0.6\textwidth]{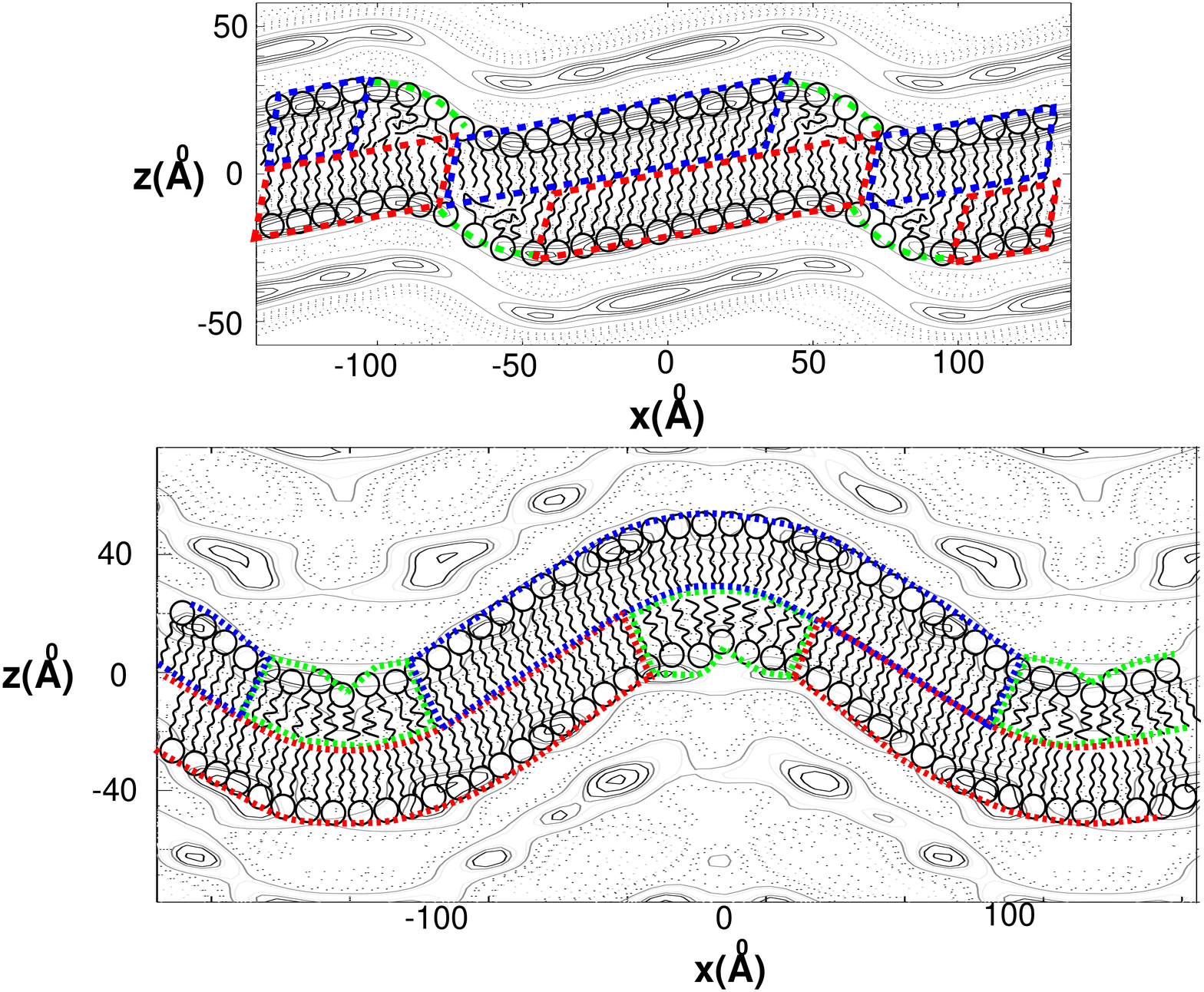} 
}
\caption{\label{fig:edm} 
Structure proposition for ripple states superimposed onto EDMs
from Ref.\ \citeonline{SRK03}. Top: Asymmetric ripple,
Bottom: Symmetric ripple. From Ref. \citeonline{LS07}.
}
\end{figure}

It is worth noting that the asymmetric ripple state does not have a
structure involving two distinct monolayers. In this respect, it differs 
fundamentally from the structures of the two neighbor phases, 
the $L_{\alpha}$ and the $L_{\beta}$ phase, and also from 
cartoons of the ripple-phase that are typically found in 
textbooks\cite{book_G}. On the other hand, a similar structure has 
recently been observed in an atomistic simulation of lecithin 
bilayers\cite{VYM05}. Our coarse-grained simulations indicate 
that this structure is generic, lipids do not need to have special 
properties to produce it. In particular, they do not need to be chiral. 
The asymmetric ripple state is closely related to the structure of the 
symmetric ripple state (which had not yet been observed in simulations). 
Our results prompted us to propose a new structural model for the 
two ripple states, which is shown in Fig.\ \ref{fig:edm}, superimposed 
onto experimental electron density maps (EDM) by Sengupta \etal\cite{SRK03} 

After the publication of our results, other authors found
similar structures with a different generic model\cite{SG08}.
It should be noted, however, that not all generic models 
produce them. A rather different modulated
phase which is less compatible with the EDMs has been reported 
from simulations of a coarse-grained membrane with soft 
DPD interactions\cite{KLS04}. Packing effects hence seem to play 
a role in stabilizing the specific structure(s) of the experimental
ripple phase. Indeed, our simulations indicate that one major player 
is the {\em splay} of the lipids within the monolayers\cite{LS07}, 
\ie the variations of the local nematic order, which is sustained 
by packing effects.

\vspace*{\baselineskip}

{\bf Interactions with membrane proteins}

Based on this successful membrane model, we proceeded to
study interactions of membranes with nanosized inclusions.
The idea is, of course, that these inclusions would be generic 
models for transmembrane (integral) proteins. Membrane-protein 
interactions have been the subject of considerable theoretical 
work in the past decades, often based on elastic approaches. 
The work described below aimed at testing such theories, 
rather than describing a particular protein. We considered 
the infinite-dilution limit, where proteins are sparse in the
membrane. At finite concentrations, integral proteins may
alter the membrane properties ({\em e.g.}, membrane thinning 
\cite{H06,LS06,PDP09}), which will in turn influence the
membran-protein interactions.

Many theories for inclusion-membrane interactions were worked 
out for the case of cylindrical inclusions, \ie cylinders 
with orientation in the direction of the bilayer normal.
They are taken to represent simple helical proteins, such
as gramicidin. One frequently addressed question relates to
the distortion (compression/expansion) of membranes in the vicinity 
of hydrophobic cylinders that are longer or shorter than the
the membrane thickness. Such 'hydrophobic mismatch' supposedly
leads to membrane-induced interactions between inclusions, which 
may induce clustering\cite{K98}. The effect has been verified 
experimentally with systematic studies of gramicidin\cite{HHW99} 
and synthetic model peptides\cite{SB02,PK03}. We shall not
review the huge amount of theoretical literature here. A list 
of references with brief discussion is given in Ref.\ \citeonline{WBS09}.

We have modeled our ''protein'' in two different ways -- as straight, 
infinitely long cylinder with a hydrophobic section of given length $L$ 
(corresponding to the situation usually studied in theoretical work), 
and as freely rotating hydrophobic cylinder of finite length $L$ with 
hydrophilic semi-spherical caps at both ends. The diameter of the 
cylinder was chosen $1.8$nm such that it roughly matches that of a 
$\beta$-helix (\eg gramicidin). We should mention that other 
simulation studies of hydrophobic mismatch interactions between 
cylindrical inclusions have been published very recently
\cite{MVS08,SGW08}, which however did not include the quantitative 
comparison with theory on which our work focusses.

\begin{figure}[t]
\centerline{
\includegraphics[width=0.48\textwidth]{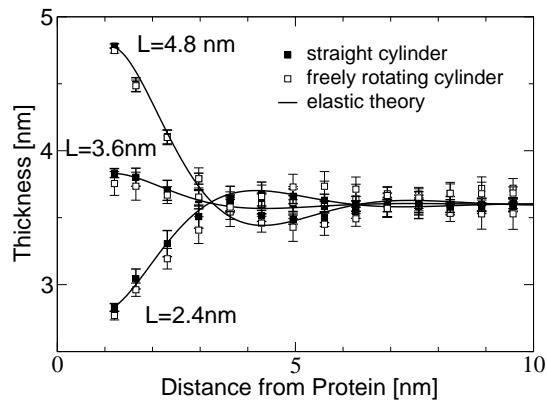} 
}
\caption{\label{fig:profiles} Thickness profiles around inclusion
for straight cylinders (closed symbols) and freely rotating
spherocylinders (open symbols) compared with elastic theory, for 
hydrophobically matched inclusion ($L=3.6$nm) and inclusions with positive
and negative hydrophobic mismatch ($L=4.8$nm and $L=2.4$nm,
respectively). After Ref. \citeonline{WBS09}.
}
\end{figure}

As main reference theory, we use here the elastic model that has served
us so well to account for the fluctuations of pure bilayers, supplemented 
with boundary conditions on the bilayer thickness and the curvature of 
the thickness profile at the surface inclusion\cite{BB06,BB07}
(see Ref.\ \citeonline{WBS09} for a physical motivation of this particular choice 
of boundary conditions). First we consider the bilayer thickness distortion 
around a single inclusion. Fig.\ \ref{fig:profiles} shows some corresponding 
radial profiles. The results for the two protein models are almost identical, 
and they can be fitted very nicely by the elastic theory.

\begin{figure}[t]
\centerline{
\includegraphics[width=0.49\textwidth]{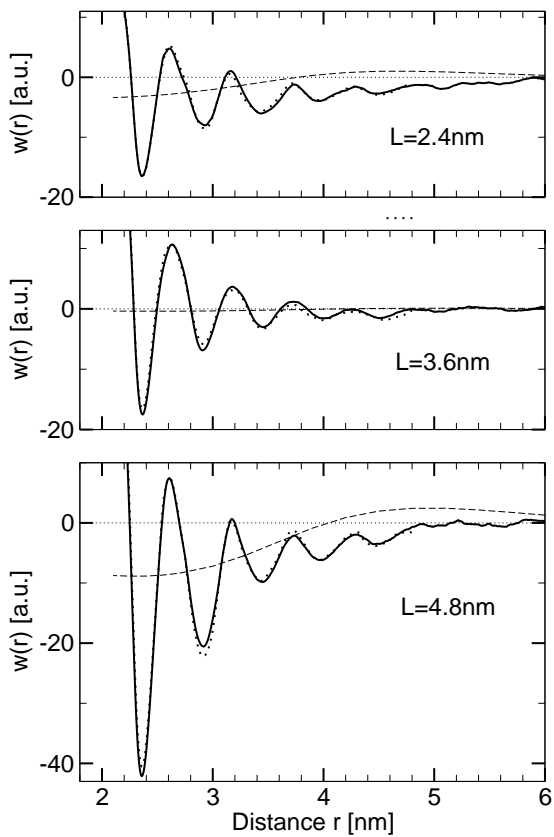}  
}
\caption{\label{fig:pmf} Potential of mean force (pmf) $w(r)$
between inclusions as a function of inclusion distance for straight
cylinders (closed lines) and freely rotating spherocylinders 
(dotted lines). Thin dashed lines show the prediction of the elastic 
theory. After Refs.\citeonline{WBS09,W08}.
}
\end{figure}

Next we examine the potential of mean force (pmf) $w(r)$ between two inclusions.
It is given by $w(r) = -\kB \ln g(r)$ with the Boltzmann factor $\kB$ and the 
pair correlation function $g(r)$. The latter has been determined from 
simulations of a system containing two inclusions by a combination of 
umbrella sampling and reweighting methods. Some results are shown in 
Fig.\ \ref{fig:pmf}. As before, the curves for the two protein models are 
almost identical, except at very close distances where the two proteins 
are in direct contact (the corresponding data are outside of the range
of the figure). The pmf has an oscillatory component, which can be explained 
by lipid packing effects. For hydrophobically mismatched inclusions, this 
packing interaction is superimposed by an additional smooth attractive interaction,
which we identify with the hydrophobic mismatch interaction. At small
protein distances, the shape of the latter is compatible with the 
prediction of the elastic theory. At larger distances, the theory predicts 
a weak oscillatory behavior (with a wavelength much larger than that
of the packing interaction), which is not observed in the simulations.
The oscillations in the theory can be traced back to a soft peristaltic 
membrane mode (see the peak of $\langle |t(q)|^2 \rangle$ in 
Fig.\ \ref{fig:fluctuations}), which also leaves a clear oscillatory
signature in the thickness profiles (Fig.\ \ref{fig:profiles}, see also 
Refs.\ \citeonline{VSS05,CP07}). Apparently, the effect of this mode on 
the lipid-mediated interactions is destroyed.

\vspace*{\baselineskip}

{\large \bf 4. Conclusions and Outlook}

Studying amphiphilic systems with generic models has a long-standing tradition.
We hope that our brief review has given a taste of the power of this approach.
It is not only highly valuable for investigating large-scale properties of
amphiphilic systems, which have been the traditional target of generic modeling, 
it can also be used to address open questions regarding the properties of membranes 
on molecular scales.

Specifically, we have discussed the use of generic models to study
membrane structure and membrane phase transitions, the statics and
kinetics of self-assembly, and also to test the validity of 
continuum theories against simulations of (coarse-grained) molecular
systems. This last aspect is particularly important because it bridges 
between generic models that represent different levels of 
coarse-graining. Continuum theories are usually constructed
heuristically based on symmetry considerations, and guided
by the idea that they should be as simple as possible. They are
genuinely generic at the continuum level. Generic molecular 
models can be used to test their validity and their limitations.
Bridging between different generic levels could eventually result
in a ``generic multiscale approach'', where generic models 
at different coarse-graining levels are used concertedly
to study non-specific properties and universal processes 
in amphiphilic systems or other complex materials.

The work presented here has resulted from enjoyable common work
with my students and postdocs Claire Loison, Xuehao He, Olaf Lenz, 
and Beate West, and from fruitful collaborations with Kurt Kremer, 
Michel Mareschal, J\"org Neder, Peter Nielaba, and Frank Brown. 
It was funded by the german science foundation (DFG) within
the SFB 613 and by the Humboldt foundation. The computer simulations
were carried out at the computer centers of the Max-Planck society
(Garching), the computing center of the Commisariat a l'Energie Atomique
(Grenoble), the John-von Neumann computing center (J\"ulich),
and at the center for parallel computing PC${}^2$ (Paderborn).

\renewcommand{\refname}{}

\end{document}